\documentclass[%
reprint,
showpacs,
showkeys,
preprintnumbers,
nofootinbib,
amsmath,amssymb,aps,prb]{revtex4-1}
\usepackage{graphics}
\usepackage{epsfig}
\usepackage{dcolumn}
\usepackage{color}
\usepackage{bm}
\usepackage{subfigure}
\usepackage{hyperref}
\usepackage[mathlines]{lineno}

\begin{document}


\title{Effects of short-range electron-electron interactions in doped graphene}



\author{Faluke Aikebaier}
\affiliation{%
 Department of Physics and Electrical Engineering, Linnaeus University, Kalmar, Sweden
}%
\author{Anna Pertsova}%
\affiliation{%
 Department of Physics and Electrical Engineering, Linnaeus University, Kalmar, Sweden
}%
\author{Carlo M. Canali}%
\affiliation{%
 Department of Physics and Electrical Engineering, Linnaeus University, Kalmar, Sweden
}%



\begin{abstract}
We study theoretically the effects of short-range electron-electron interactions on 
 the electronic structure of graphene, in the presence of single substitutional impurities. 
 Our computational approach is based on 
 the $\pi$ orbital tight-binding approximation for graphene, with the 
 electron-electron interactions treated self-consistently at the level of the mean-field Hubbard model. 
  We compare explicitly non-interacting and interacting cases with varying interaction strength and 
   impurity potential strength. 
 We focus in particular on the interaction-induced modifications 
  in the local density of states around the impurity, which is a quantity that 
  can be directly probed by scanning tunneling spectroscopy of doped graphene. 
  We find that the resonant character of the impurity states 
  near the Fermi level is enhanced by the interactions.     
 Furthermore, the size of the energy gap, which opens up at 
 high-symmetry points of the Brillouin zone of the supercell upon doping,    
  is significantly affected by the interactions. The details of this effect 
   depend subtly on the supercell geometry. 
  We use a perturbative model to explain these features and find quantitative 
  agreement with numerical results.
\end{abstract}

\maketitle

\section{\label{sec:level1}Introduction}
Graphene -- a two-dimensional allotrope of carbon, has attracted considerable attention in recent years, 
 largely due to its remarkable electronic properties stemming from the massless-Dirac-fermion nature of 
 its low-energy quasiparticle states.~\cite{2009RvMP...81..109C} 
 Most of the electronic properties of graphene that have been studied 
experimentally can be well described by non-interacting single-particle theory.  
 However, electron-electron interactions in graphene are expected to be strong. 
  In undoped clean graphene, the density of states 
  at the Fermi level vanishes and therefore the Coulomb potential 
  is not screened.~\cite{2007PhT....60h..35G,RevModPhys.84.1067}  
 Recent experiments have shown that unscreened Coulomb interactions lead to reshaping of 
the ideal conical energy dispersion expected in graphene.~\cite{2011NatPh...7..701E} More precisely, the Fermi 
 velocity near the Dirac point acquires a logarithmic correction as a result 
  of interactions.  
 
 From a theoretical viewpoint, it can be shown that this logarithmic enhancement arises from 
the non-local exchange interaction, 
 already at the level of the first-order Hartree-Fock perturbation theory.~\cite{katsnelson2012graphene} 
  Hence, it is the long-range nature of the electron-electron interactions in graphene  
  that is responsible for the logarithmic correction, the most striking interaction 
  effect observed so far in this material in the absence of external magnetic fields. 
 As a result, theoretical work has mostly focused on investigations of long-range interactions in graphene, 
using a variety of techniques ranging from mean field~\cite{PhysRevB.84.085446} to renormalization group 
approaches.~\cite{Gonzalez1994595, PhysRevB.59.R2474} 
  
 It should be noted, however, that there are several important conditions that need to be 
 satisfied in order to observe significant long-range interaction effects 
experimentally.  It is necessary to 
 be able to probe a wide range of carrier concentrations and 
 to tune the Fermi level sufficiently close the Dirac point, 
  where the renormalization of the Fermi velocity is expected to be dramatic due to 
  the vanishing density of states. Moreover, spurious screening effects, e.g. 
  dielectric screening from the substrate, should be avoided.  
This makes undoped high-quality suspended graphene 
an ideal platform for studying the effects 
of long-range electron-electron interactions.~\cite{2011NatPh...7..701E}
 On the contrary, in the case of graphene on a substrate 
 or in the presence of disorder and impurities these 
 effects are less relevant. 
 
 In particular, doping introduces a finite density of states at the Dirac 
 point of graphene and the long-range part of 
 the Coulomb potential is screened. In this case, 
 short-range interactions become crucial. If these interactions 
 are fairly strong, they can lead to interesting effects 
  on the electronic structure, especially 
   on the impurity states in the vicinity of the Fermi level. 
 In fact, estimates of the on-site Hubbard $U$ parameter 
  in carbon-based molecules~\cite{Parr1950,PhysRevLett.56.1509} suggest that 
   the short-range Coulomb interactions 
   among $\pi$-electron in graphene can be indeed quite large, i.e. 
    of the order of $10$~eV. A similar value is obtained 
    by accurate \textit{ab initio} calculations.~\cite{PhysRevLett.106.236805}

In this paper, we study the effects of short-range electron-electron interactions 
 on the electronic structure of doped graphene. We should note 
 that the importance of impurity effects in graphene has been addressed in many theoretical 
 studies.~\cite{PhysRevB.73.241402,PhysRevB.86.045448,PhysRevB.80.214201,PhysRevLett.92.256805,PhysRevB.73.125414,
PhysRevB.84.245446,PhysRevB.85.035404} A number of 
 interesting features have been revealed, including the opening of the 
  gap upon doping~\cite{PhysRevB.80.214201,PhysRevB.86.045448} and the appearance of impurity (acceptor or donor) states in the vicinity of 
  the Fermi level.~\cite{PhysRevB.73.241402,PhysRevB.84.245446} There is also a great interest in 
   addressing these properties experimentally~\cite{Zhao19082011,doi:10.1021/nn1002425,PhysRevB.85.161408} 
   since doping graphene with impurities is one way to further explore and tune its 
   electronic, magnetic and transport properties. 
However, the interplay of short-range electron-electron interactions 
and impurity potentials in graphene has not yet been fully explored.
 
We use a single-band ($\pi$ orbital) tight-binding (TB) model to describe the electronic structure 
 of graphene. A supercell method is employed to 
  study the effects of finite doping. A substitutional impurity 
 is introduced in the TB Hamiltonian as a local modification of the on-site 
potential at the impurity site. Here we focus on attractive impurity 
potentials, mimicking nitrogen impurity atoms which are typical 
 dopants in graphene.~\cite{Zhao19082011} The short-range interactions are described by means of the 
 Hubbard model in the mean-field approximation, which is the simplest way  of  
treating the many-body interacting problem. 
 Interaction terms are introduced at each site in the TB Hamiltonian. 
 We use a numerical self-consistent scheme to 
 account for a redistribution of electronic charge 
 around the impurity caused by interactions. As the output of our numerical 
 calculations, we 
 obtain the band structure and the local density of states (LDOS) around the impurity site. 
  Furthermore, we calculate scanning tunneling microscopy (STM) images 
   by integrating the LDOS over a small energy window above the Fermi level. 

By using this approach, we show that 
 short-range interactions introduce several remarkable 
  features in the electronic structure of doped graphene. 
Importantly, they enhance the resonant character of states 
localized in real space around the impurity, which are induced in the vicinity of the Dirac point. The complex
 interplay between short-range interactions and impurity potential is also responsible for non-trivial gaps at
 high-symmetry crossing points in the band structure of graphene, in particular at the Dirac point.

The paper is organized as follows. In Sec.~\ref{sec:level2} we introduce our TB model and 
  describe how the impurity potential and short-range interactions are incorporated in the Hamiltonian.  
  We also provide some details of the self-consistent supercell calculations.   
  Our findings are described in Sec.~\ref{sec:level3}. In particular, 
  in Sec.~\ref{bands} we focus on the effects of interactions on the band structure of graphene 
and on some issues related to the  supercell geometry. In Sec.~\ref{ldos} we discuss the changes in 
 the resonant character of the LDOS around the impurity for varying impurity potential strength and interaction 
strength. The comparison between the simulated STM topographies 
for non-interacting and interacting cases is provided. Finally, we draw some conclusions.

\section{\label{sec:level2}Methodology}

The second-quantized Hamiltonian for interacting electrons on a honeycomb lattice in the presence 
of impurity can be written as  
\begin{equation}\label{eq:HamiltonianForTheSystem}
H={\sum_{i{{\sigma}}}{\varepsilon}_{i}c^{\dagger}_{i{\sigma}}c_{i{\sigma}}}+{\sum_{\left\langle 
{i,j}\right\rangle{\sigma}}}{t_{ij} c^{\dagger}_{i{\sigma}}c_{j{\sigma}}}+U_{\mathrm{im}}{c^{\dagger}_{0{\sigma}}c_{0{\sigma}}}+U{\sum_{i}}{n_{i{\uparrow}}n_{i{\downarrow}}}.
\end{equation}
Here $c^{\dagger}_{i{\sigma}}$ and $c_{i{\sigma}}$ are the creation and annihilation operators 
for electron on site $i$ and with spin $\sigma$; 
${\varepsilon}_{i}$ and $t_{ij}$ are on-site energies and hopping parameters, respectively. Only  
 hopping between nearest neighbors on the honeycomb lattice is included. We 
 assume that the TB parameters are uniform, except for the 
  on-site energy at the impurity site, and we use the values 
  obtained by fitting the TB band structures to density functional theory calculations, 
  namely $\varepsilon_{i}=0$ and $t_{ij}=-2.97$~eV.~\cite{PhysRevB.66.035412}  

  The third term in Eq.~(\ref{eq:HamiltonianForTheSystem}) represents the local 
impurity potential, with $U_{\mathrm{im}}$ being the impurity potential strength 
 ($U_{\mathrm{im}}<0$ for attractive impurity). In our calculations we 
 use $U_{\mathrm{im}}=-10$~eV and $U_{\mathrm{im}}=-20$~eV in order 
 to obtain visible trends for the impurity states in the vicinity of the Fermi level. 
  
 The last term describes 
 the on-site interaction between two electrons with opposite spins on site $i$ (including the impurity site), 
 with $U$ ($U\ge 0$) being the Hubbard $U$ parameter, which expresses the strength of the 
 intra-atomic Coulomb repulsion. Here $n_{i{\sigma}}$ is the number operator, 
 defined as $n_{i{\sigma}}=c^{\dagger}_{i{\sigma}}c_{i{\sigma}}$. We consider 
  $U=0$, or non-interacting case, and $U=9.3$~eV, 
  which is the value obtained for graphene using the constrained Random Phase Approximation method.~\cite{PhysRevLett.106.236805} 
   In order to extract the trends in the electronic structure with 
   increasing the interaction strength we also use a larger value of $U=20$~eV. 
  
In the mean-field approximation, the two-body interaction term in Eq.~(\ref{eq:HamiltonianForTheSystem}) becomes 
\begin{equation}\label{eq:HamiltonianInMeanField}
U{\sum_{i}}{n_{i{\uparrow}}n_{i{\downarrow}}}=
U{\sum_{i}}{\left({\left\langle {n_{i{\downarrow}}}\right\rangle}c^{\dagger}_{i{\uparrow}}c_{i{\uparrow}}
+{\left\langle {n_{i{\uparrow}}}\right\rangle}c^{\dagger}_{i{\downarrow}}c_{i{\downarrow}}\right)},
\end{equation}
where $\left\langle {n_{i{\sigma}}}\right\rangle$ is the average electron occupation number, or density, for 
spin-up  ($\sigma=\uparrow$)  and spin-down ($\sigma=\downarrow$) electrons. Here we consider 
  a non spin-polarized case so that ${\left\langle {n_{i{\uparrow}}}\right\rangle}={\left\langle {n_{i{\downarrow}}}\right\rangle}$.

In pristine graphene with the Fermi level exactly at the Dirac point, the average electron occupation number is a constant 
 equal to $1/2$. Adding a mean-field field on-site potential does 
  not break the translational invariance of the crystal   
  and the average occupation number remains constant. In fact, in orthogonal basis such a potential merely introduces 
   a rigid shift of the energy bands (note that in non-orthogonal basis  
 the interplay between the overlap integrals and the on-site interactions  
  can lead to renormalization of the Fermi velocity~\cite{unpbl}). 
 
 However, the presence of both mean-field on-site interactions and 
  impurity potential can lead to non-trivial effects in the electronic 
  structure. In this case, 
   the potential at each site 
depends on the average occupation number ${\left\langle {n_{i{\sigma}}}\right\rangle}$, which is not 
necessarily the same on all sites. As a result, when a carbon atom is replaced by an impurity, 
there will be a redistribution of electronic charge in the system. In order to capture this effect, 
we need to perform self-consistent calculations for the Hamiltonian in Eq.~(\ref{eq:HamiltonianForTheSystem})
 and (\ref{eq:HamiltonianInMeanField}). 

At each step of the self-consistent cycle,  
the average occupation number for site $i$ is calculated as
\begin{equation}
\left\langle {n_{i{\sigma}}}\right\rangle={\frac{1}{N}}{\sum_{k}^{occ}}{\left|{b^{k}_{i{\sigma}}}\right|^{2}},
\end{equation}
where $N$ is the number of $k$-points in the Brillouin zone and $b^{k}_{i{\sigma}}$ are 
 the coefficients in the expansions of the wavefunctions of the Hamiltonian in terms of 
  the localized atomic orbitals $\left| i\sigma \right\rangle$.  These 
   are obtained by diagonalization of the Hamiltonian at each $k$-point. 
   The sum 
   runs over all occupied states up to the Fermi level. Note that all calculations 
   are done at half-filling. 
  
As initial values we use the occupation numbers calculated for a non-interacting problem, 
i.e. for a supercell of graphene with impurity ($U_{\mathrm{im}}\ne 0$) and with $U=0$.  
The criterion of self-consistency is
\begin{equation}
{\sum_{i{\sigma}}}\left|{{ \left\langle n_{i{\sigma}}\right\rangle }^{s}-{\left\langle n_{i{\sigma}}\right\rangle}^{s-1}}\right|<{\eta},
\end{equation}
where $s$ is the index of the self-consistent cycle and ${\eta}$ is a small parameter (we choose ${\eta}=10^{-7}$). 
  We use a linear mixing scheme, in which the input  density 
 $\left\langle n_{i\sigma} \right\rangle^{s+1}_{\mathrm{in}}$ at step  $s+1$ 
 is calculated as a linear combination of outputs $\left\langle n_{i\sigma} \right\rangle^{s}_{\mathrm{out}}$ and $\left\langle n_{i\sigma} \right\rangle^{s-1}_{\mathrm{out}}$  
from two  previous steps
\begin{equation}
\left\langle n_{i\sigma} \right\rangle^{s+1}_{\mathrm{in}}=\left({1-{\lambda}}\right)\left\langle n_{i\sigma} \right\rangle^{s-1}_{\mathrm{out}}+
{\lambda}\left\langle n_{i\sigma} \right\rangle^{s}_{\mathrm{out}},
\end{equation} 
where ${\lambda}$ is the mixing coefficient; we use ${\lambda}=0.25$, which 
allows us to achieve self-consistency in less than $100$ steps.
   
\begin{figure}
\centering
\includegraphics[width=0.97\linewidth]{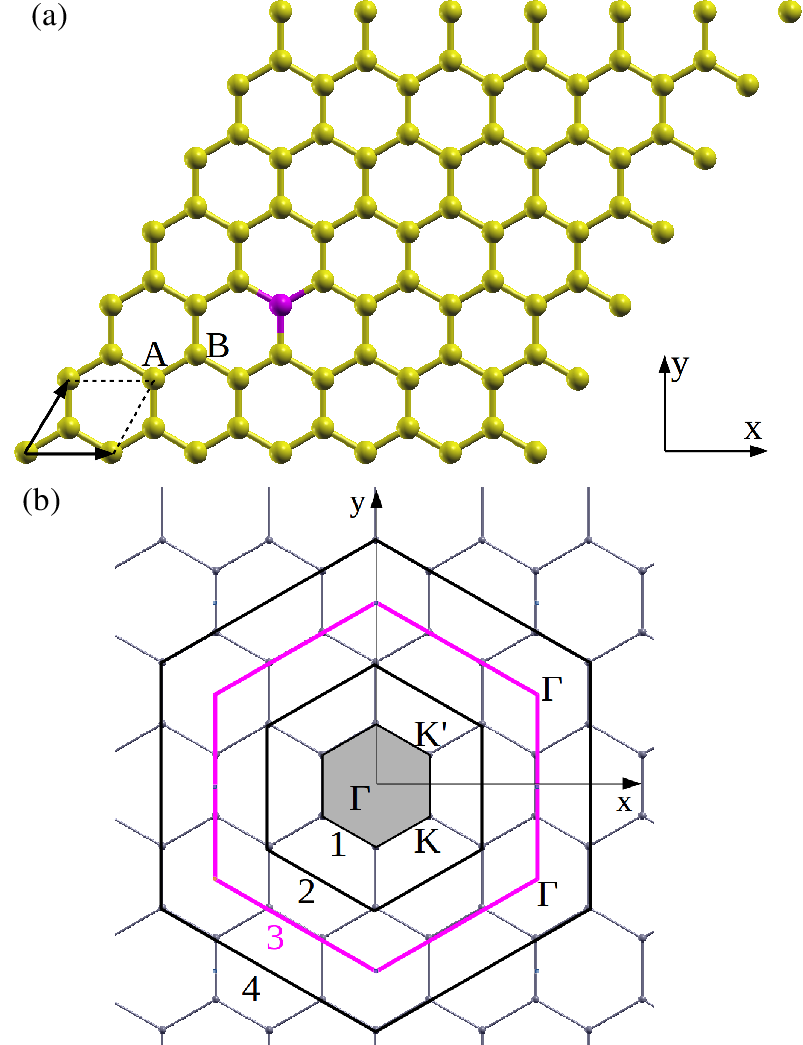}
\caption{(Color online) (a) $6\times6$ supercell of graphene with a substitutional impurity. A magenta sphere 
represents the impurity atom. Dashed lines mark the unit cell of pristine graphene and 
 arrows show the primitive lattice vectors. A and B denote carbon atoms in the two equivalent sublattices.  
 (b) Brillouin zone folding in graphene. Shaded area (1) represents the first Brillouin zone 
 of a $p\times p$ supercell of graphene. Numbered curves correspond to the first Brillouin zone of the 
  unit cell for different $p$: $p=3q-1$ (2), $p=3q$ (3) and $p=3q+1$. For $p=3q$ the Dirac points of graphene ($K$, $K^{\prime}$) 
  are mapped onto $\Gamma$ point of the folded Brillouin zone. In other cases, $K$ and $K^{\prime}$ 
  are mapped onto  $K$ and $K^{\prime}$ of the folded Brillouin zone.}
\label{fig:00} 
\end{figure}

In order to 
model the effect of finite impurity concentration, 
 we construct a $p\times p$ supercell 
  by replicating a graphene unit cell $p$ time along each of the two-dimensional lattice vectors 
 [see Fig.~\ref{fig:00}(a)]. 
  The impurity atom substitutes a carbon atom in the supercell.  
  In this work, we use two different supercells with $p=6$ and $p=7$.   
  Atomic concentration of the dopants depends 
on the size of the supercell so the concentration is slightly different for the two choices, 
namely 1.0$\%$ for a $7\times 7$ and 1.4$\%$ for a $6\times 6$ supercell. 
 It is known that for $p=3 q$, where $q$ is an integer, the Dirac points of graphene, $K$ and $K^{\prime}$, 
  are mapped onto the $\Gamma$ point of the Brillouin zone of the supercell.~\cite{0957-4484-24-22-225705,PhysRevB.86.045448,PhysRevB.81.245420} 
   as illustrated in Fig.~\ref{fig:00}(b). 
  This does not happen 
  if $p$ is not divisible by $3$. Therefore, 
  the $6\times 6$ supercell is special. As we explain in Sec.~\ref{bands}, the effects of impurity potential 
  and interactions in this case are rather non-trivial. 
 This is the main reason for considering two different supercell sizes. 

\section{\label{sec:level3}Results}

\subsection{\label{bands}Bandstructure}

It is known that the finite amount of doping opens up 
an energy gap at the Dirac point of graphene.~\cite{PhysRevB.86.045448,PhysRevLett.92.256805,PhysRevB.84.245446} 
Here we address the question of how the details of the bandstructure near the gap are affected by interactions. 

\begin{figure}
\centering
\includegraphics[width=0.97\linewidth]{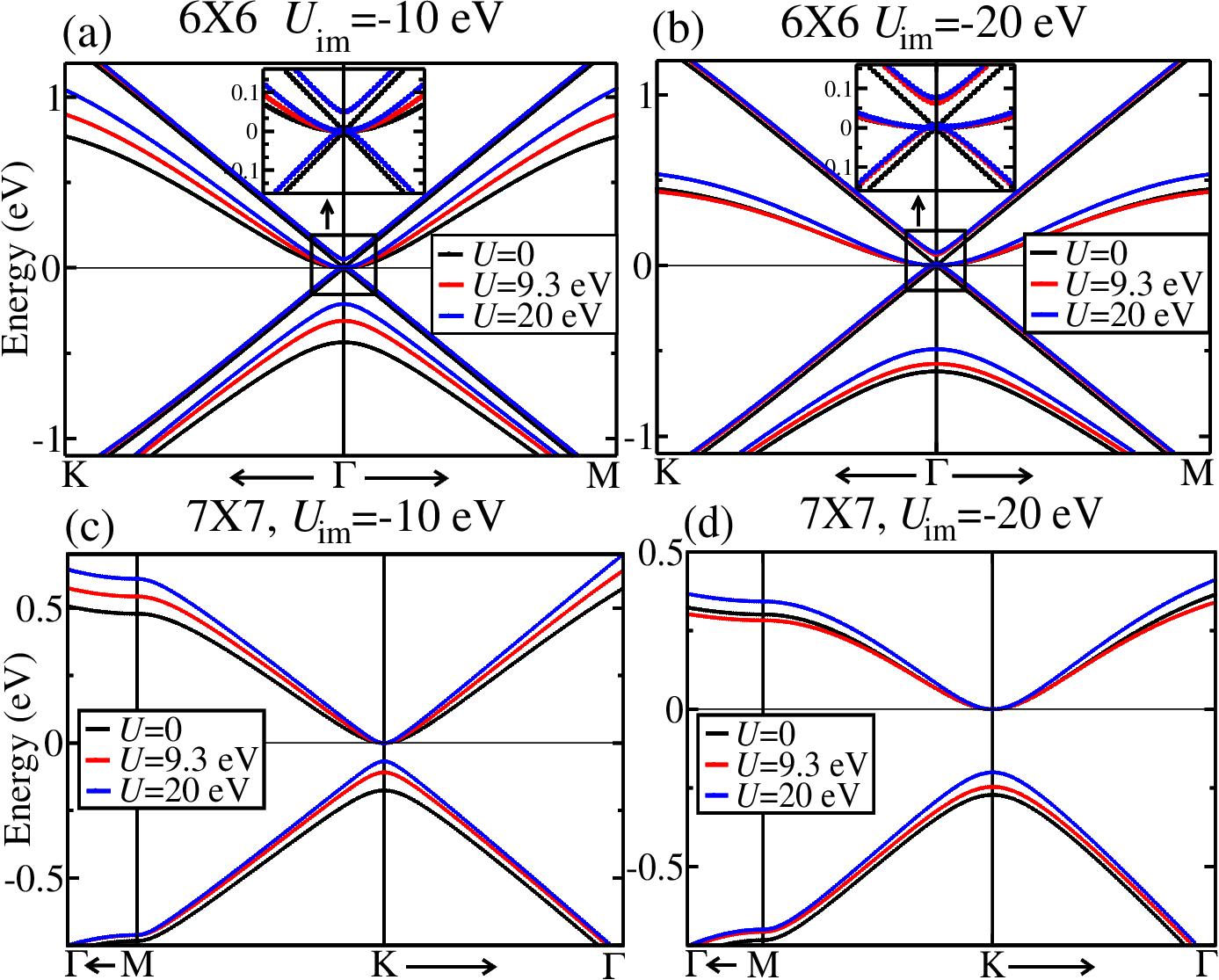}
\caption{(Color online) Bandstructure of doped graphene along the high-symmetry lines of the  
 Brillouin zone for $6\times 6$ (a,b) and $7\times 7$ (c,d) supercell, for 
 varying impurity potential strength $U_{\mathrm{im}}$ and interaction strength $U$. Left panels 
 are for  $U_{\mathrm{im}}=-10$~eV, right panels for $U_{\mathrm{im}}=-20$~eV. In each panel 
 three cases are shown: $U=0$ (black), $U=9.3$~eV (red) and $U=20$~eV (blue). Horizontal line 
  in (a) and (b) is the position of the 
  doubly degenerate state at $\Gamma$, adjusted to the value found 
   for $U=0$ (see text for details). Horizontal line in (c) and (d) marks the 
   conduction band maximum, which has been aligned with the value found for  $U=0$.}
\label{fig:02} 
\end{figure}

We start with a special supercell geometry $p\times p$, with $p$ divisible by $3$ ($p=6$ in our calculations). 
Figure~\ref{fig:02}(a)-(b) shows the bandstructure of the $6\times 6$ supercell with impurity potential $U_{\mathrm{im}}=-10$~eV  
 and $U_{\mathrm{im}}=-20$~eV, respectively, for three values of the interaction strength, $U=0$, 
  $U=9.3$~eV and $U=20$~eV. 
  Note that in the bandstructure calculations, different impurity potential and interaction strength 
   introduce a shift of the energy bands with respect to a reference case, i.e.  non-interacting 
   pristine graphene  ($U_{\mathrm{im}}=0$ and $U=0$). In order to examine the features around the gap 
    for different choices of parameters, we align the position of the doubly degenerate 
    state (see the discussion below) in all curves in Fig.~\ref{fig:02}(a)-(b) to the value found for $U=0$ for a given impurity potential strength. 
  
As we mentioned before, in the case of $p=6$ both $K$ and $K^{\prime}$ are mapped 
 onto ${\Gamma}$ point,~\cite{0957-4484-24-22-225705,PhysRevB.86.045448,PhysRevB.81.245420} 
  producing four degenerate states at $\Gamma$ in the absence of impurities 
  and interactions. When one carbon atom in the supercell is substituted 
  by an impurity atom, a gap opens up between two states at $\Gamma$, however the other two states remain degenerate.  
 More precisely, for $U=0$ three of the four states at $\Gamma$ are degenerate while one of the states moves away from the Dirac point. 
  This situation is referred to as the \textit{pseudogap}~\cite{PhysRevB.86.045448}  since there 
  is still a pair of linearly dispersed states crossing at the neutrality point. Hence, effectively there is no band gap 
   for this special supercell size. 
  
One can clearly see from Fig.~\ref{fig:02}(a) 
that the size of the pseudogap decreases with increasing the interaction strength. 
 On-site interactions cause a redistribution of charge around the impurity. We 
 find that in the case of the attractive impurity potential $U_{\mathrm{im}}=-10$~eV,  
 the total average occupation at the impurity site decreases from $\left\langle n_{0} \right\rangle=0.89$ 
 in the non-interacting case to $\left\langle n_{0} \right\rangle=0.82$ for $U=9.3$~eV. 
 The on-site Coulomb repulsion prevents extra charge from accumulating on the impurity and 
 therefore the strength of the impurity potential is effectively reduced. 
For the impurity potential which is twice stronger,  the pseudogap 
 for the same values of $U$ is noticeably larger [see Fig.~\ref{fig:02}(b)]. 
 
In addition to a large pseudogap, for $U\ne 0$ there is also a smaller pseudogap which opens up at $\Gamma$ [see the insets in Fig.~\ref{fig:02}(a)-(b)]. 
 There are now only two generate states at $\Gamma$, while the other two states shift, respectively, above and below the crossing point. 
 Interestingly, the effect of interactions on the small pseudogap 
  is opposite to that on the large one, e.g. its value increases with increasing the interaction strength. 
  A perturbative model described in Appendix~\ref{sec:appendix} suggests that the 
 smaller gap results from the contribution of the states localized on 
 sublattice B, if we assume that the impurity is substituted in sublattice A. 
 Within our model, this effect is solely due to interactions (the contribution of sublattice B to the gap is zero in the absence of interactions~\cite{PhysRevB.86.045448}).
 We find the following values for the two gaps at $\Gamma $ from analytical calculations (see Appendix \ref{sec:appendix} for details). 
 In the non-interacting case and $U_{\mathrm{im}}=-10$~eV, the large pseudogap is -0.56~eV. For $U=9.3$~eV, the large pseudogap decreases   
to 0.39~eV. At the same time, a small pseudogap of 0.05~eV opens up. For $U=20$~eV, the large and small pseudogaps 
become 0.28~eV and 0.06~eV, respectively. 
These values are all in good agreement with the pseudogaps found in Fig.~\ref{fig:02}(a). 
 Note that for the larger impurity potential, the trends with increasing $U$ are the same, however for a given $U$ both gaps are larger 
  than in $U_{\mathrm{im}}=-10$~eV case. 
 Analytical calculations using the perturbative model in this case also agree with numerical results.

Similar features are found for a regular $7\times 7$ supercell [see Fig.~\ref{fig:02}(c) and (d)]. Note that the 
conduction band minima at $\Gamma$ have been aligned with the reference $U=0$ case. The main difference from the 
$6\times 6$ supercell is that in this case there 
is a real band gap at $K$ ($K^\prime$). Our perturbative analysis shows that 
 there is no contribution from sublattice B to the gap at the Dirac point, if we assumed that the impurity 
 is substituted in sublattice A. As in the case of the $6\times 6$ supercell, the size of the gap decreases 
  with increasing the interaction strength. Analytically, for $U_{\mathrm{im}}=-10$~eV we find a gap of 0.20 eV for
  $U=0$, 0.14 eV for $U=9.3$~eV and 0.10 eV for $U=20$~eV. These values are in good agreement with 
  numerical calculations. Somewhat smaller values of the gaps compared to a $6\times 6$ supercell 
   for the same $U_{\mathrm{im}}$ and $U$ are expected since the atomic concentration of impurities 
   is smaller. 

Bandstructure calculations presented in this section lead to a conclusion that short-range interactions effectively reduce the strength 
 of the impurity potential, which results in a decrease of the large gap (pseudogap) at the Dirac point.  
 In oder to see how the character, e.g. the energy and the spatial extent, of the electronic states around the Dirac point 
  is affected by interactions, we need to look at the LDOS around the impurity.

\subsection{\label{ldos}Local density of states}

Calculations of LDOS at the impurity site reveal several important features. 
 A substitutional impurity introduces electronic states at energies 
  comparable to the impurity potential ($|U_\mathrm{im}|\sim 10$~eV), i.e. 
  far away from the Fermi level. However, there are also states 
   appearing in the vicinity (within $\sim 1$~eV) of the Fermi level.~\cite{PhysRevB.77.115109,PhysRevB.86.045448,
   PhysRevB.73.241402,PhysRevB.75.125425} 
    These states  are the  
    most relevant for the low-energy electronic properties 
    of graphene and will be examined in detail. 
  
\begin{figure}
\centering
\includegraphics[width=0.97\linewidth]{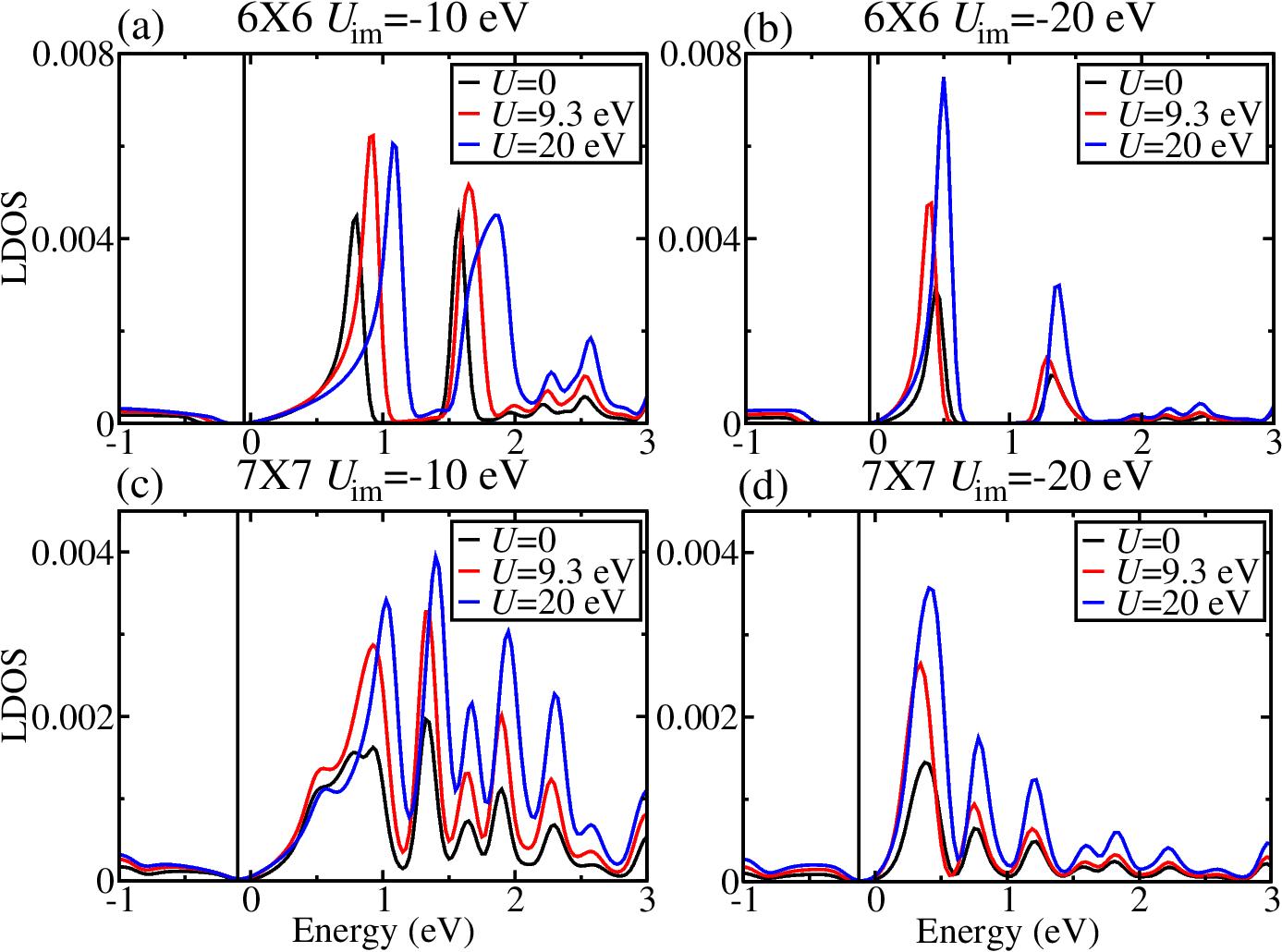}
\caption{(Color online)  LDOS of doped graphene at the impurity site for   
  $6\times 6$ (a,b) and $7\times 7$ (c,d) supercell, for 
 varying impurity potential strength $U_{\mathrm{im}}$ and interaction strength $U$. Left panels 
 are for  $U_{\mathrm{im}}=-10$~eV, right panels for $U_{\mathrm{im}}=-20$~eV. In each panel 
 three cases are shown: $U=0$ (black), $U=9.3$~eV (red) and $U=20$~eV (blue). 
 Vertical lines mark the position of the Fermi level (see text for details).}
\label{fig:01} 
\end{figure}

  Figure~\ref{fig:01} shows the double- or multi-peak impurity resonances 
close to the Fermi level for the $6\times 6$ 
and $7\times 7$ supercells, respectively, for different impurity potential and interaction strengths. 
    The multi-peak structure of the impurity resonances most likely originate 
 from the long-range interaction, or interference,  between the impurity potentials, caused by the periodicity of the supercell 
geometry.~\cite{PhysRevB.86.045448} 

With increasing the impurity potential strength, 
the resonant peaks move closer to low energies. 
This is very similar to the case of  strong potential impurities on the surface of a topological insulator.~\cite{PhysRevB.85.121103} As shown 
in Fig.~\ref{fig:01}(a) and (b) with $U=0$, the double-peak resonance in the $6\times 6$ supercell approaches the Fermi level as 
  $U_{\mathrm{im}}$ increases. At the same time 
  its amplitude decreases.
 This finding is in  
agreement with semi-analytical calculations in 
Refs.~\onlinecite{PhysRevB.77.115109,PhysRevB.73.241402,PhysRevB.75.125425}. 
The effect of the impurity potential is more 
striking in the case of a regular $7\times 7$ supercell 
[Fig.~\ref{fig:01}(c) and (d)]. 
 In this case, in addition 
  to shifting the peaks to lower energies, 
  a stronger impurity potential $U_{\mathrm{im}}=-20$~eV also 
  makes the peaks narrower, thus enhancing 
   their resonant character [see in particular the first peak in 
   Fig.~\ref{fig:01}(c) and (d) with $U=0$].
  
In the interacting case, the amplitude of the resonances increases with $U$.  
 This can be seen in all panels in Fig.~\ref{fig:01} with $U=9.3$~eV  and $U=20$~eV, 
 with the exception of the $6\times 6$ supercell with $U_{\mathrm{im}}=-10$~eV and $U=20$~eV, where 
 the amplitude of the peak decreases slightly. 
 In the case of $U_{\mathrm{im}}=-10$~eV, for both supercells  
 the impurity resonances move further away from the Fermi level 
 with increasing $U$. This is perfectly consistent with our observations for 
 the non-interacting case with decreasing impurity potential. 
  However, in the case of a very large impurity potential $U_{\mathrm{im}}=-20$~eV, 
 the trend in the position of the resonances is less obvious. 
 The peaks either do not move appreciably as in the case of $U=20$~eV or 
  even seem to move slightly towards the low-energy region for $U=9.3$~eV. 
 Below we elaborate more on these findings.
 
 Increasing $U$ reduces the overall strength of the impurity potential, 
 which is confirmed by decrease of the energy gap at the Dirac point due to 
 the presence of impurities (Sec.~\ref{bands}).  
However, short-range electron-electron interactions controlled by $U$ 
 do not only change the potential directly at the impurity site but 
 also affect the on-site potential and the charge density around the impurity (primarily nearest and 
  next-nearest neighbors of the impurity atom). 
 Hence, both the amplitude and the spatial extent of the impurity potential is 
 modified by interactions.  
Let us  assume that an attractive  
 impurity can be described by a delta-function potential well. 
 When interactions are included, the shape of 
  the impurity potential is smoothed out (it acquires, say, a Gaussian shape). 
Therefore, although the strength of the potential 
 is reduced by a certain amount with increasing $U$, 
 the potential can become more long-ranged (in a certain parameter space). 
 This, in turn, will increase the overlap of the potentials from neighboring cells 
 and enhance the inter-supercell interaction.
 
 This seems to be the situation for $U_{\mathrm{im}}=-20\ \text{eV}$ and 
  $U=9.3$~eV, for both choices of the supercell. 
  In this case, the impurity potential decreases slightly due to interactions 
   ($U<U_\mathrm{im}$), leading to a small decrease of the gap at the Dirac point   
 compared to $U=0$ case [Fig.~\ref{fig:02}(b) and (d)]. 
 At the same time, we find a significant difference between the average occupation numbers 
of the nearest and next-nearest neighbors for $U=0$ and $U=9.3$~eV    
 (this can be also partly seen in the STM images in Fig.~\ref{fig:04}(a)-(b) and (d)-(e), for 
 states in a small energy window above the Fermi level, where 
 the neighbors of the impurity site appear brighter in the $U=9.3$~eV case compared to $U=0$). 
 As a result, the amplitude of impurity resonances in LDOS increases but their position 
  shifts to lower energies.  
  
  In contrast to this, for  $U_{\mathrm{im}}=-20\ \text{eV}$ and $U=20$~eV, 
the average occupation numbers of the nearest and next-nearest neighbors of the impurity site do not change 
  appreciably compared to $U=0$ case. The strength of the impurity potential for this value of $U$ 
  is significantly reduced, leading to 
    a large decrease of the energy gap [Fig.~\ref{fig:02}(b) and (d)]. As a result, 
    the amplitude of impurity resonances increases further, however their position  
     remain close to the $U=0$ value. 
    These features strongly suggest that the position of the impurity resonances is sensitive to the 
    spatial extent of the impurity potential, namely the resonances move closer to zero energy 
     when the potential becomes more long-ranged.
   
\begin{figure}
\centering
\includegraphics[width=0.97\linewidth]{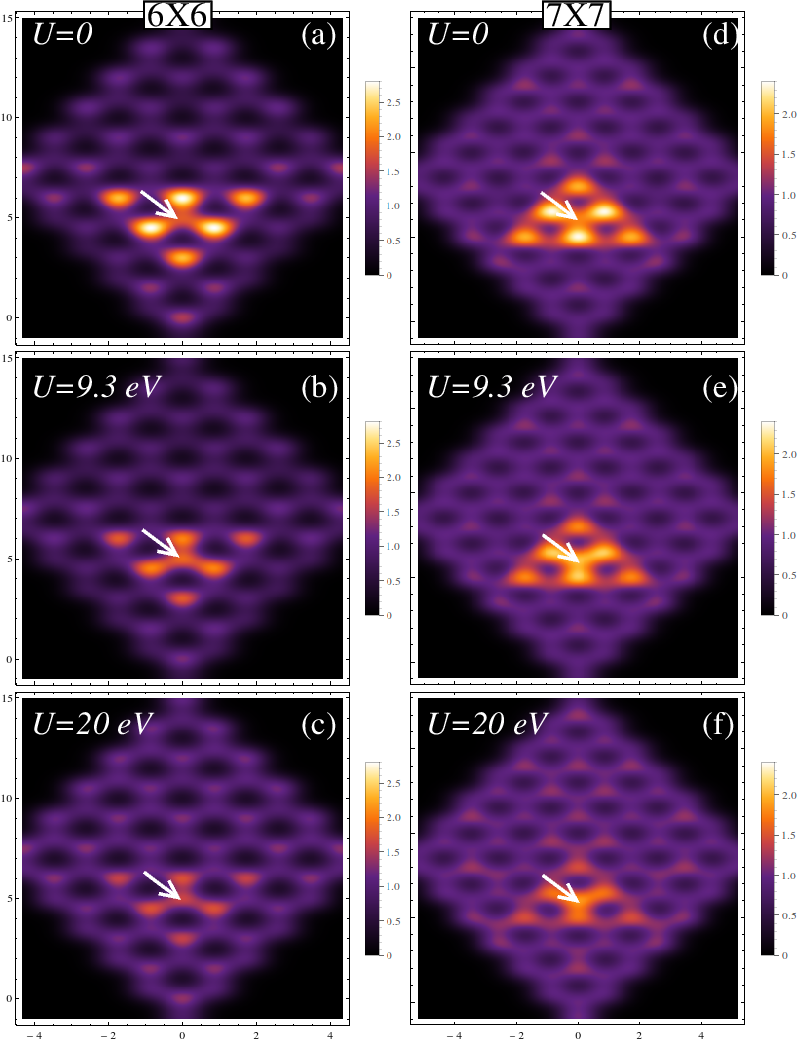}
\caption{(Color online) Simulated STM topographies (LDOS for all atoms in the supercell) 
 for $6\times 6$ (left panels) and $7\times 7$ (right panels) supercell, for 
 a fixed  impurity potential strength $U_{\mathrm{im}}=-10$~eV and varying 
 interaction strength: $U=0$ (a,d), $U=9.3$~eV (b,e) and $U=20$~eV (c,f). Arrows 
 mark the position of the impurity atom. Note the logarithmic color scale.
 }
\label{fig:03} 
\end{figure}

\begin{figure}
\centering
\includegraphics[width=0.97\linewidth]{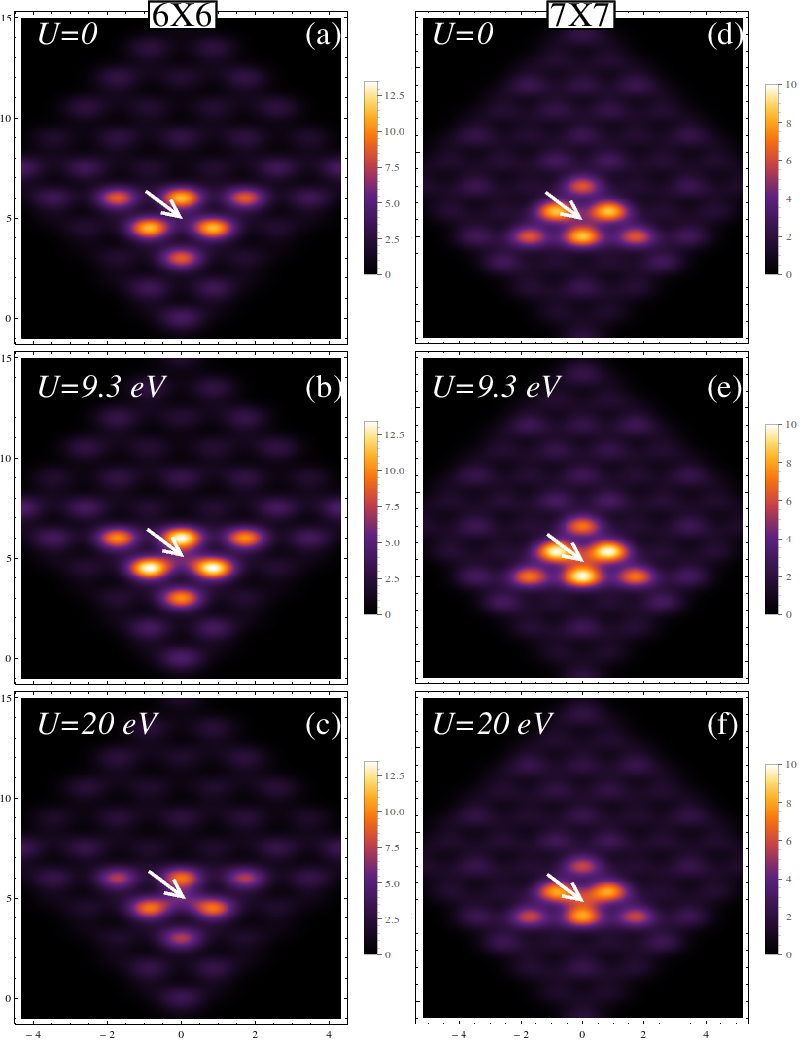}
\caption{(Color online) Same as Fig.~\ref{fig:03} but 
 with impurity potential strength $U_{\mathrm{im}}=-20$~eV.}
\label{fig:04} 
\end{figure}

To further clarify the 
  changes in the intensity and the spatial character of the low-energy impurity peaks, 
   brought about by interactions, 
  we  present the simulated STM topographies in Fig.~\ref{fig:03} and Fig.~\ref{fig:04}  
   for $U_\mathrm{im}=-10$~eV and $U_\mathrm{im}=-20$~eV, respectively. For this 
we plot LDOS for each atom in the supercell,~\footnote{The discrete values of LDOS 
 at lattice points are smeared out by adding a Gaussian broadening of 
 ${\gamma}=N_{\text{atoms}}/E_{\text{mesh}}$, where $N_{\text{atoms}}$ is 
the number of atoms in the supercell and $E_{\text{mesh}}$ is the energy mesh.
} integrated over the energy window $\Delta E$ 
above the Fermi level (we choose $\Delta E=0.25$~eV). This gives an estimate of the tunneling current 
as electrons tunnel out of the occupied states of the STM tip into the unoccupied states of  graphene. 
The increase of the electronic density of states in 
this energy window gives rise to a bright triangular feature around the impurity. 
Note that for the $6\times 6$ supercell the impurity is located in sublattice A, while 
 for the $7\times 7$ supercell it is located in sublattice B. Therefore the bright triangular features 
  in the two supercells appear rotated by $180^{\circ}$ with respect to each other.
 
The difference between non-interacting and interacting cases is clearly visible in the STM images. 
In all cases considered, the impurity site becomes progressively brighter 
 compared to its neighbors with increasing $U$.  
 This means that the electronic states 
 in the small energy window above the Fermi level become 
 more localized on the impurity site as a result of interactions. 
 This is most evident in the case of the $7\times 7$ supercell [Fig.~\ref{fig:03}(d)-(f)].  
 For $U_{\mathrm{im}}=-10$~eV, the overall 
  intensity of the images decreases with $U$. 
  For a stronger $U_{\mathrm{im}}=-20$~eV impurity potential, the trend is similar with the exception 
 of the intermediate interaction strength $U=9.3$~eV. In this case, the contribution of the nearest  
 and next-nearest neighbors is even stronger than in the $U=0$ case. This correlates  
 with the corresponding features in the LDOS discussed above.

\section{\label{sec:level5}Conclusions}

We have presented a theoretical study of the effects of 
electron-electron interactions on the electronic states of graphene in the presence of 
 substitutional impurities. 
Using a self-consistent TB model with on-site interactions treated at the mean-field level, 
we have shown that the size of the gap, which opens up at the Dirac point in graphene upon doping, 
and the character of the low-energy electronic states 
 are modified by interactions. The mechanism 
for these effects is provided by the interplay between the impurity potential and the on-site repulsion, 
which leads to significant re-arrangement of the electronic charge around the impurity 
compared to the non-interacting case.

In particular, we found that the size of the gap 
decreases with increasing the interaction strength. 
Intuitively, this can be understood as follows. In the case of 
an attractive impurity potential, which mimics nitrogen dopants in graphene, 
adding the on-site Coulomb repulsion effectively reduces the strength of the potential, 
 i.e. the depth of the potential well decreases. This is due to the fact the on-site repulsion 
 prevents extra electronic charge from accumulating on the impurity site. 
 
For a special supercell size $p\times p$, where $p$ is divisible by $3$, both 
$K$ and $K^{\prime}$ are mapped onto $\Gamma$ point of the folded Brillouin zone. 
Therefore, in the case 
of undoped non-interacting graphene, there are four degenerate states at the neutrality point. 
It is known that when the impurity potential is included, a gap (pseudogap) opens up between two of these states 
 while the other pair remains degenerate.~\cite{PhysRevB.86.045448} 
 We have shown that the size of this pseudogap 
 is reduced by interactions. Interestingly, in addition to this, 
 a small gap opens up between the second pair 
 of states at the $\Gamma$ point, which are otherwise degenerate in the absence of 
 interactions. We explain these features both qualitatively and quantitatively, 
 using a perturbative model based on the generalization of the approach developed by Lambin 
 \textit{et al.}~\cite{PhysRevB.86.045448} to the interacting case.  
 
Furthermore, we have studied the behavior of the impurity-induced electronic states with and without interactions. 
There are two groups of states which can be detected in the density of states when a carbon atom in 
 the supercell is replaced by an impurity atom. First, there are states which emerge far away from  
 the Fermi energy, with their energies of the same order of magnitude as the impurity potential (
 $\approx 10$~eV in our calculations). Second, there are states appearing close to the Fermi energy 
 and are therefore of particular interest. Although the 
  way the electron-electron interactions affect the LDOS at low energies 
  in general depends on the impurity concentration, we found   
   clear trends in the behavior of the impurity-resonances as both parameters, 
   i.e. the interaction strength and the impurity potential strength,  
  are modified.  
  
  Regardless of the interactions, the impurity levels 
   move closer to the low-energy region (e.g. to the original Dirac point) 
   with increasing the impurity potential strength.  
   This finding is consistent with previous 
   calculations for graphene.~\cite{PhysRevB.77.115109,PhysRevB.73.241402,PhysRevB.75.125425} 
  Similar result was found in another 
   class of Dirac materials, namely in three-dimensional topological insulators   
   in the presence of strong potential impurities on the surface.~\cite{PhysRevB.81.233405,PhysRevB.85.121103}
 However, our self-consistent calculations for graphene in the presence of both impurities and interactions
  reveal novel features. 
 For a fixed impurity potential,   
 sort-range interactions 
tend to enhance the amplitude of the impurity-resonances in the vicinity 
of the Fermi level. The position of the resonances is also affected by 
the spatial extent of the effective impurity potential, which is modified by interactions. 
As the interaction strength increases, the states become more localized on the impurity atom. 
 The differences in the spatial distribution of the low-energy impurity states 
  in the non-interacting and interacting cases are clearly detectable in the simulated STM topographies. 

\appendix
\section{Band gap in doped graphene supercell in the presence of interactions}
\label{sec:appendix}

We use a simple perturbative model, based on the model proposed in Ref.~\citenum{PhysRevB.86.045448} for non-interacting graphene, 
 to  explain how the 
impurity potential and electron-electron interactions affect the band 
structure of doped graphene.

We first review this model for the non-interacting case with impurity.~\cite{PhysRevB.86.045448} The Hamiltonian for such
 a system can be written in Dirac's notation as 
 \begin{equation}\label{eq:HamiltonianForNonInSingleImpuirty}
H=H_0+H_1,
\end{equation}
where $H_0$ is the Hamiltonian of non-interacting pristine graphene 
\begin{equation}\label{eq:HamiltonianForNonInSingleImpuirty1}
H_0={\sum_{u}}{\left|{u}\right\rangle E_{u}\left\langle {u}\right|}+{\sum_{u,v}}{\left|{u}\right\rangle t_{uv}\left\langle {v}\right|}, 
\end{equation}
and $H_1$ is the perturbation introduced by the periodic arrangement of impurities
\begin{equation}\label{eq:HamiltonianForNonInSingleImpuirty2}
H_1=\sum_{u\in 1}\left|{u}\right\rangle U_{\mathrm{im}}\left\langle {u}\right|.
\end{equation}
Here $\left|u\right\rangle$ is the atomic orbital associated with site $u$ ($\left\langle u|u^{\prime}\right\rangle=\delta_{u u^{\prime}}$), 
$E_u$ and $t_{uv}$ are the on-site energies and the nearest-neighbors hopping integrals, 
respectively;  $U_{\mathrm{im}}$ is the impurity potential and 
 the sum over $u\in 1$ refers to all impurity atoms belonging to sublattice 1, which substitute one carbon atom in each  $p\times p$ supercell .

In pristine graphene [Eq.~(\ref{eq:HamiltonianForNonInSingleImpuirty1})], there are four states with zero energy, 
two for the two sublattices and two for the non-equivalent points $K$ and $K^{\prime}$ in the Brillouin zone (we omit the spin indices 
for simplicity). The corresponding Bloch functions 
can be written as 
\begin{equation}\label{eq:BlochFunction}
\left|K^{\mathrm{A(B)}}\right\rangle={\frac{1}{\sqrt{N_{\mathrm{A(B)}}}}}{\sum_{u{\in}\mathrm{A(B)}}}{\text{e}^{i\mathbf{K}{\cdot}\mathbf{u}}\left|u\right\rangle},
\end{equation}
where $\mathbf{K}$ is a vector in the reciprocal space corresponding to either $K$ or $K^{\prime}$, $\mathbf{u}$ is the position of 
site $u$ in real space, and $N_{\mathrm{A(B)}}$ is the number of atoms in sublattice A(B).

The task is to calculate the first order corrections to the energy states at 
 the Dirac point due to the impurity potential, by using degenerate state perturbation theory. 
 Let us assume that the impurity is substituted in sublattice A.   
Then the states $\left|K^\mathrm{B}\right\rangle$ and $\left|K^{\prime \mathrm{B}}\right\rangle$ 
 have zero amplitudes on atoms in sublattice A and these states are eigenstates of zero energy. 
 Therefore,  we only need to consider the subspace of degenerate eigenstates formed by the states $\left|K^\mathrm{A}\right\rangle$ and $\left|K^{\prime\mathrm{A}}\right\rangle$. 
 We use Eq.~(\ref{eq:HamiltonianForNonInSingleImpuirty2}) and Eq.~(\ref{eq:BlochFunction}) to calculate the following matrix elements 
\begin{eqnarray}\label{eq:MatrixElementA1NonIn}
V_{11}&=&\left\langle K^{\mathrm{A}}\right|H_{1}\left|K^\mathrm{A}\right\rangle = {\frac{U_\mathrm{im}}{N_\mathrm{A}}}={\frac{U_\mathrm{im}}{p^{2}}},\\
V_{22}&=&\left\langle K^{\prime\mathrm{A}}\right|H_{1}\left|K^{\prime\mathrm{A}}\right\rangle =V_{11}= {\frac{U_\mathrm{im}}{p^{2}}},
\end{eqnarray}
\begin{eqnarray}\label{eq:MatrixElementA2NonIn}
V_{12}&=&\left\langle K^\mathrm{A}\right|H_{1}\left|K^{\prime\mathrm{A}}\right\rangle={\frac{1}{N_\mathrm{A}}}\sum_{u\in 1}e^{i\left(\mathbf{K}^{\prime}-\mathbf{K}\right){\cdot}\mathbf{u}}\nonumber\\
&=&{\frac{U_\mathrm{im}}{p^{2}}}{\delta}_{\mathbf{K}^{\prime}-\mathbf{K},\mathbf{G}},\\
V_{21}&=&\left\langle K^{\prime\mathrm{A}}\right|H_{1}\left|K^\mathrm{A}\right\rangle=V_{12}={\frac{U_\mathrm{im}}{p^{2}}}{\delta}_{\mathbf{K}^{\prime}-\mathbf{K},\mathbf{G}},
\end{eqnarray}
where $N_\mathrm{A}=p^{2}$ since for $p\times p$ supercell, we have $N=2p^{2}$ atoms and $N_\mathrm{A}=N_\mathrm{B}=p^{2}$ atoms in  each sublattice. 
$V_{12}(V_{21})$ in Eq.~(\ref{eq:MatrixElementA2NonIn}) is not zero only when the vector $\mathbf{K}^{\prime}-\mathbf{K}$ is equal to the reciprocal 
lattice vector $\mathbf{G}$. 
This condition is satisfied only if $p$ is divisible by 3, i.e. if $p=3q$, where $p$, $q$ are integers. 
This is essentially the result obtained in Ref.~\onlinecite{PhysRevB.86.045448} and it is confirmed by our 
calculations presented in Fig.~\ref{fig:02} for $U=0$ case. 

The first corder corrections $E^{(1)}$ to the energy states are then given by the eigenvalues of matrix $V$, with matrix elements 
 $V_{ij}$ ($i,j=1,2$) defined above. This gives $E^{(1)}=0$ and $E^{(1)}=2 U_\mathrm{im}/p^2$ if $p=3q$, and $E^{(1)}=U_\mathrm{im}/p^2$ if $p\ne 3q$. 
 Hence, in the case when $p$ is divisible by 3, all four states are mapped onto $\Gamma$ point. 
 Three of these states, two of which are localized on sublattice B and one on sublattice A, have zero energy. The remaining state is shifted down in energy 
  by $2 U_\mathrm{im}/p^2$ ($U_\mathrm{im} < 0$ for attractive impurity), producing a pseudogap of magnitude $2 U_\mathrm{im}/p^2$. 
  This is exactly the situation shown in  Fig.~\ref{fig:02}(a) and (b) for the $6\times 6$ supercell. 
 In the case when $p$ is not divisible by 3, the degeneracy between $K$ and $K^{\prime }$ is lifted and there are now two states at each of these points, 
 separated by a gap of $U_\mathrm{im}/p^2$. This is the situation for the $7\times 7$ supercell in Fig.~\ref{fig:02}(c) and (d).
 
Combining the results for $p=3q$ and $p\ne 3q$, to the lowest order in the perturbation theory 
 the gap (pseudogap) induced by the impurity potential can be written as  
\begin{equation}\label{eq:GapForNonInteractingCase}
E_{\mathrm{gap}}={\frac{U_{\mathrm{im}}}{p^{2}}}+{\frac{U_\mathrm{im}}{p^{2}}}{\delta}_{\mathbf{K}^{\prime}-\mathbf{K},\mathbf{G}}.
\end{equation}
For  $U_\mathrm{im}=-10$~eV, $E_\mathrm{gap}=-0.56$~eV if $p=6$ and $E_\mathrm{gap}=-0.2$~eV if $p=7$. These values coincide with the gaps found numerically 
[see Fig.~\ref{fig:02}(a) and (c) with $U=0$].

Below we  extend this model to the interacting case. 
 The Hamiltonian of the interacting system can be written as
 \begin{equation}\label{eq:HamiltonianForInSingleImpuirty}
 H=H_0+H_1+H_2,
\end{equation}
where $H_0$ and $H_1$ are given by Eq.~(\ref{eq:HamiltonianForNonInSingleImpuirty1}) and Eq.~(\ref{eq:HamiltonianForNonInSingleImpuirty2}), 
respectively, and $H_2$ is the perturbation introduced by short-range interactions which is given by 
 \begin{equation}\label{eq:HamiltonianForInSingleImpuirty1}
 H_2={\sum_{u}}{\left|u\right\rangle\left(U{\cdot}\left\langle n_{u{\sigma}}\right\rangle\right)\left\langle u\right|}.
\end{equation}
In accordance with Eq.~(\ref{eq:HamiltonianInMeanField}), $\left\langle n_{u{\sigma}}\right\rangle$ 
 is the average electron occupation number on site $u$  corresponding to spin $\sigma$. 
 When  the mean-field interaction term is included, the energy bands acquire a rigid shift. In order to compare 
the results to the non-interacting case, we need to subtract this shift. 
Let us assume that it is proportional to an average quantity $\left\langle {\bar{n}_{u{\sigma}}}\right\rangle$, which is a constant. Then the energy that needs to 
 be subtracted is $U{\cdot}\left\langle {\bar{n}_{u{\sigma}}}\right\rangle$. 

We now calculate the first-order corrections to the energy states at the Dirac point due to interactions using the same procedure. 
Since the sum in Eq.~(\ref{eq:HamiltonianForInSingleImpuirty1}) 
  can be decomposed into the sum over $u\in\mathrm A$ and the sum over $u\in\mathrm{B}$, we can calculate corrections due to these two 
  terms separately. For each of them we need to consider either the subspace formed by the states  $K^\mathrm{A}$ and $K^\mathrm{\prime A}$, 
  or by the states  $K^\mathrm{B}$ and $K^\mathrm{\prime B}$. Let us assume for simplicity that all occupation numbers 
  for atoms in sublattice A are approximately the same and equal 
 to $\left\langle \bar{n}_{u{\sigma}}\right\rangle$,  
  except for the impurity site. Then the corresponding matrix elements are give by 
 \begin{eqnarray}\label{eq:eq:MatrixElementA1In}
W^\mathrm{A}_{11}&=&\left\langle K^\mathrm{A}\right|H_{2}\left|K^\mathrm{A}\right\rangle=
{\frac{1}{2N_\mathrm{A}}}{\sum_{u\in\mathrm{A},{\sigma}}}{U\left({\left\langle n_{u{\sigma}}\right\rangle-\left\langle \bar{n}_{u{\sigma}}\right\rangle}\right)}\nonumber\\
&=&{\frac{U}{p^{2}}}\left({\left\langle n_{u^{\prime}{\sigma}}\right\rangle-\left\langle \bar{n}_{u{\sigma}}\right\rangle}\right),\\
%
W^\mathrm{A}_{12}&=&\left\langle K^\mathrm{A}\right|H_{2}\left|K^{\prime \mathrm{A}}\right\rangle\nonumber\\
&=&{\frac{U}{p^{2}}}\left(\left\langle n_{u{^\prime}{\sigma}}\right\rangle-\left\langle \bar{n}_{{u\sigma}}\right\rangle\right){\delta}_{\mathbf{K}^{\prime}-\mathbf{K},\mathbf{G}},\\
W^\mathrm{A}_{22}&=&\left\langle K^\mathrm{\prime A}\right|H_{2}\left|K^\mathrm{\prime A}\right\rangle=W^\mathrm{A}_{11},\\
W^\mathrm{A}_{21}&=&\left\langle K^\mathrm{\prime A}\right|H_{2}\left|K^{\mathrm{A}}\right\rangle=W^\mathrm{A}_{12},
\end{eqnarray}
where $u^{\prime}$ is the impurity site;  2 in the denominator stands for spin. As before, first order 
corrections to the energy states are given by the eigenvalues of matrix $W^{\mathrm{A}}$, with matrix elements $W^\mathrm{A}_{ij}$ ($i,j=1,2$), 
and differ for $p=3q$ and $p\ne 3q$. These corrections give the following contribution to the energy gap at the Dirac point
\begin{equation}\label{eq:GapForInteractingCaseA}
E^{\mathrm{int}(\mathrm{A})}_{\mathrm{gap}}={\frac{U}{p^{2}}}\left({\left\langle n_{u^{\prime}{\sigma}}\right\rangle-
\left\langle \bar{n}_{u{\sigma}}\right\rangle}\right)\left({1+{\delta}
_{\mathbf{K}^{\prime}-\mathbf{K},\mathbf{G}}}\right).
\end{equation}

In a similar way, we calculate the first-order corrections in the subspace formed by states $K^\mathrm{B}$ and $K^\mathrm{\prime B}$. The corresponding 
 matrix elements are given by
\begin{eqnarray}\label{eq:eq:MatrixElementB1In}
W^\mathrm{B}_{11}&=&\left\langle K^\mathrm{B}\right|H_{2}\left|K^\mathrm{B}\right\rangle=
{\frac{1}{2N_\mathrm{B}}}{\sum_{u\in\mathrm{B},{\sigma}}}{U\left({\left\langle n_{u{\sigma}}\right\rangle-\left\langle \bar{n}_{u{\sigma}}\right\rangle}\right)}\nonumber\\
&=&{\frac{U}{p^{2}}}\sum_{u\in\text{nn of }u^{\prime}}\left({\left\langle n_{u{\sigma}}\right\rangle-\left\langle \bar{n}_{u{\sigma}}\right\rangle}\right),\\
%
W^\mathrm{B}_{12}&=&\left\langle K^\mathrm{B}\right|H_{2}\left|K^{\prime\mathrm{B}}\right\rangle\nonumber\\
&=&{\frac{U}{p^{2}}}\sum_{u\in\text{nn of }u^{\prime}}\left(\left\langle n_{u{\sigma}}\right\rangle-\left\langle \bar{n}_{u{\sigma}}\right\rangle\right){\delta}_{\mathbf{K}^{\prime}-\mathbf{K},\mathbf{G}},\\
W^\mathrm{B}_{22}&=&\left\langle K^\mathrm{\prime B}\right|H_{2}\left|K^\mathrm{\prime B}\right\rangle=W^\mathrm{B}_{11},\\
W^\mathrm{B}_{21}&=&\left\langle K^\mathrm{\prime B}\right|H_{2}\left|K^\mathrm{ B}\right\rangle=W^\mathrm{B}_{12},
\end{eqnarray} 
where we assumed that all atoms in sublattice B have  approximately the same occupation 
  $\left\langle \bar{n}_{u{\sigma}}\right\rangle$, except for the nearest neighbors of the impurity atom. This is a reasonable 
  assumption since these atoms are most strongly affected by the impurity. Then the contribution to the energy 
  gap, stemming from corrections to the states localized on sublattice B, is given by 
\begin{equation}\label{eq:GapForInteractingCaseB}
E^{\mathrm{int}(\mathrm{B})}_{\mathrm{gap}}={\frac{U}{p^{2}}}\sum_{u\in\text{nn of }u^{\prime}}\left({\left\langle n_{u{\sigma}}\right\rangle-
\left\langle \bar{n}_{u{\sigma}}\right\rangle}\right)\left({1+{\delta}
_{\mathbf{K}^{\prime}-\mathbf{K},\mathbf{G}}}\right).
\end{equation}
%

Finally, combining the corrections due to the impurity potential [Eq.~(\ref{eq:GapForNonInteractingCase})] and due to 
interactions [Eq.~(\ref{eq:GapForInteractingCaseA}) and Eq.~(\ref{eq:GapForInteractingCaseB})], we obtain the expression for the energy gap at the Dirac point
\begin{eqnarray}\label{gap_int}
 E^{\mathrm{int}}_{\mathrm{gap}}&=&
\left\lbrace \left({\frac{U_\mathrm{im}}{p^{2}}} + {\frac{U}{p^{2}}}\left({\left\langle n_{u^{\prime}{\sigma}}\right\rangle-
\left\langle \bar{n}_{u{\sigma}}\right\rangle}\right)\right)\left(1+{\delta}_{\mathbf{K}^{\prime}-\mathbf{K},\mathbf{G}}\right)\right\rbrace\nonumber\\
&+&\left\lbrace {\frac{U}{p^{2}}}\sum_{u\in\text{nn of }u^{\prime}}\left({\left\langle n_{u{\sigma}}\right\rangle-
\left\langle \bar{n}_{u{\sigma}}\right\rangle}\right)\left({1+{\delta}
_{\mathbf{K}^{\prime}-\mathbf{K},\mathbf{G}}}\right)  \right\rbrace\nonumber,\\
\end{eqnarray}
where the expression inside the first curly bracket is due to the states localized on sublattice A, while 
the second one is due the states localized on sublattice B. 

Let us now summarize what happens to the four zero energy states, when both 
the impurity potential and interactions are present. 
When $p=3q$, all four states are mapped onto $\Gamma$ point. As one can see from Eq.~(\ref{gap_int}),  
 one of the states, corresponding to sublattice A, remains at zero energy, while the other one 
 shifts by $E^{\mathrm{int}(\mathrm{A})}_{\mathrm{gap}}=2\left[\frac{U_\mathrm{im}}{p^{2}}+ 
 {\frac{U}{p^{2}}}\cdot\left( {\left\langle n_{u^{\prime}{\sigma}}\right\rangle-
\left\langle \bar{n}_{u{\sigma}}\right\rangle} \right)\right]$. We can estimate this quantity by taking the values of 
 the rigid band shift and the average occupation numbers for impurity and its nearest neighbors from our 
 numerical calculations (see Table~\ref{Table1}, numbers are similar for the two supercells). For $U_\mathrm{im}=-10$~eV, 
$U=9.3$~eV and $p=6$, this energy shift is negative and is equal to $-0.39$~eV. This corresponds to the 
large pseudogap (below the Dirac point) that we identified in Fig.~\ref{fig:02}(a) for this choice of $U$. 
 In a similar way, one of the states localized on sublattice B remains at zero energy, while the other 
 one shifts up in energy by $E^{\mathrm{int}(\mathrm{B})}_{\mathrm{gap}}=2\frac{U}{p^{2}}\sum_{u\in\text{nn of }u^{\prime}}\left({\left\langle n_{u{\sigma}}\right\rangle-
\left\langle \bar{n}_{u{\sigma}}\right\rangle}\right)=0.054$~eV. This small energy shift is identical to 
the small pseudogap (above the Dirac point) found in Fig.~\ref{fig:02}(a).

\begin {table}
\caption {Values of band shifts and average occupation numbers used in Eq.~(\ref{gap_int}) for $U_\mathrm{im}=-10$~eV and $U=9.3$~eV.} \label{Table1} 
\begin{center}
\begin{tabular}{|l|l|l|l|}
\hline
Band shift &$\left\langle \bar{n}_{u{\sigma}} \right\rangle$ & $\left\langle n_{u^{\prime}{\sigma}} \right\rangle$ (imp.)  & $\left\langle n_{u{\sigma}}\right\rangle$ (nn of imp.)  \\
\hline
 4.60~eV & 0.49 & 0.82 & 0.46 \\
\hline
\end{tabular}
\end{center}
\end{table}

When $p\ne 3$, the degeneracy between $K$ and $K^{\prime}$ is lifted. There are two states at each of these points, one localized on sublattice A and the other 
one on sublattice B. The energy gap between the states is given by $E^{\mathrm{int}}_{\mathrm{gap}}=\frac{U_\mathrm{im}}{p^{2}}+ 
 {\frac{U}{p^{2}}}\cdot\left( {\left\langle n_{u^{\prime}{\sigma}}\right\rangle-
\left\langle \bar{n}_{u{\sigma}}\right\rangle} \right)+\frac{U}{p^{2}}\sum_{u\in\text{nn of }u^{\prime}}\left({\left\langle n_{u{\sigma}}\right\rangle-
\left\langle \bar{n}_{u{\sigma}}\right\rangle}\right)$. For $U_\mathrm{im}=-10$~eV, 
$U=9.3$~eV and $p=7$, $E^{\mathrm{int}}_{\mathrm{gap}}=0.14$~eV, which is in good agreement with Fig.~\ref{fig:02}(c).

\bibliographystyle{apsrev4-1}
\bibliography{apstemplate}{}

\end{document}